\documentclass[conf]{new-aiaa}
\usepackage[utf8]{inputenc}

\usepackage{graphicx}
\usepackage{amsmath}
\usepackage[version=4]{mhchem}
\usepackage{siunitx}
\usepackage{longtable,tabularx}
\title{Pressure-Induced Separation of a Laminar Boundary Layer over a Partially-Slip Wall}

\author{Benjamin~Kellum~Cooper\footnote{Undergraduate Student}, Benjamin S. Savino\footnote{Ph.D. Candidate}, John Marshall Cooper\footnote{Undergraduate Student}, Taiho Yeom\footnote{Assistant Professor}, and Wen Wu\footnote{Assistant Professor and AIAA member. Corresponding author. Email address: wu@olemiss.edu}}
\affil{Department of Mechanical Engineering, University of Mississippi, University, MS, 38677, USA}

\begin{document}

\maketitle

\section{Introduction}
The separation of boundary layers over solid surfaces has been extensively investigated due to its great impacts on aerodynamic performance and relevance to the aviation and defense sectors~\cite{RLSimpson89,NaM98a}. 
Flow separation may occur on a surface exhibiting localized partial-slip properties. Examples include scenarios such as water film formation on airfoils and turbines in high-humidity regions and rainfall, as well as lubricated surfaces and coatings treated with strengthened hydrophobicity in engineering applications. These surfaces manifest inhomogeneous slipping conditions to the adjacent fluid flow. 
Because of the slipping surface, the velocity deficit at the wall is reduced and it is expected to postpone flow reversal and accumulation of surface vorticity. As a result, slippery surfaces have the potential to delay separation and modulate the wake. 

While numerous studies have been done on the impact of slipping surfaces on boundary layers, the role of surface slipping on flow separation is less focused. 
Legendre et al.~\cite{LegendreLM09} observed delayed separation, decreased shed vorticity, and increased vortex shedding frequency in their direct numerical simulations (DNS) of a slipping circular cylinder. 
The delayed onset of separation was confirmed by Sooraj et al.~\citep{Soorajetal20} and Muralidhar et al.~\citep{MURALIDHARetal11} in their experiments of flow past cylinders coated with superhydrophobic paint or fabric. Sooraj et al. described that the recirculation region increased when the separated shear layer was steady. They observed that the recirculation length was reduced when vortices were formed in the separated shear layer, yet Muralidhar et al. reported an increase. Li et al.~\cite{Lietal14} found that where the slip boundary was applied was important in determining the flow structure and force around the circular cylinder. Slipping boundary conditions applied around the maximum width exhibited the maximum drag reduction in their study. Ceccacci et al.~\cite{Ceccacci1etal22} performed DNS of flow separation on the lee side of a bump in a channel. They observed a delayed onset of separation and a reduced separation bubble. No vortex shedding was reported. 

In these investigations, the flow separation is associated with the configuration-dependent geometric curvature. The essential features of flow separation, however, exist without geometric curvature~\cite{NaM98a}. Therefore, we focus on the pressure-gradient-induced flow separation over a flat plate in this work.
A benchmark laminar separating boundary layer that has been well characterized for no-slip plates in the literature~\cite{HosseinverdiF19} is utilized. The goal is to investigate the effects of upstream streamwise slipping variations on the onset of separation and reattachment. A partial-slip region is assigned upstream of the separation region through boundary conditions. The changes in the separation region, especially the onset of separation and the size, are reported in this abstract. The full paper will include complete statistics of a more systematic parametric study.
\section{Methodology}
\begin{figure}
    \centering
    \includegraphics[width=0.44\textwidth]{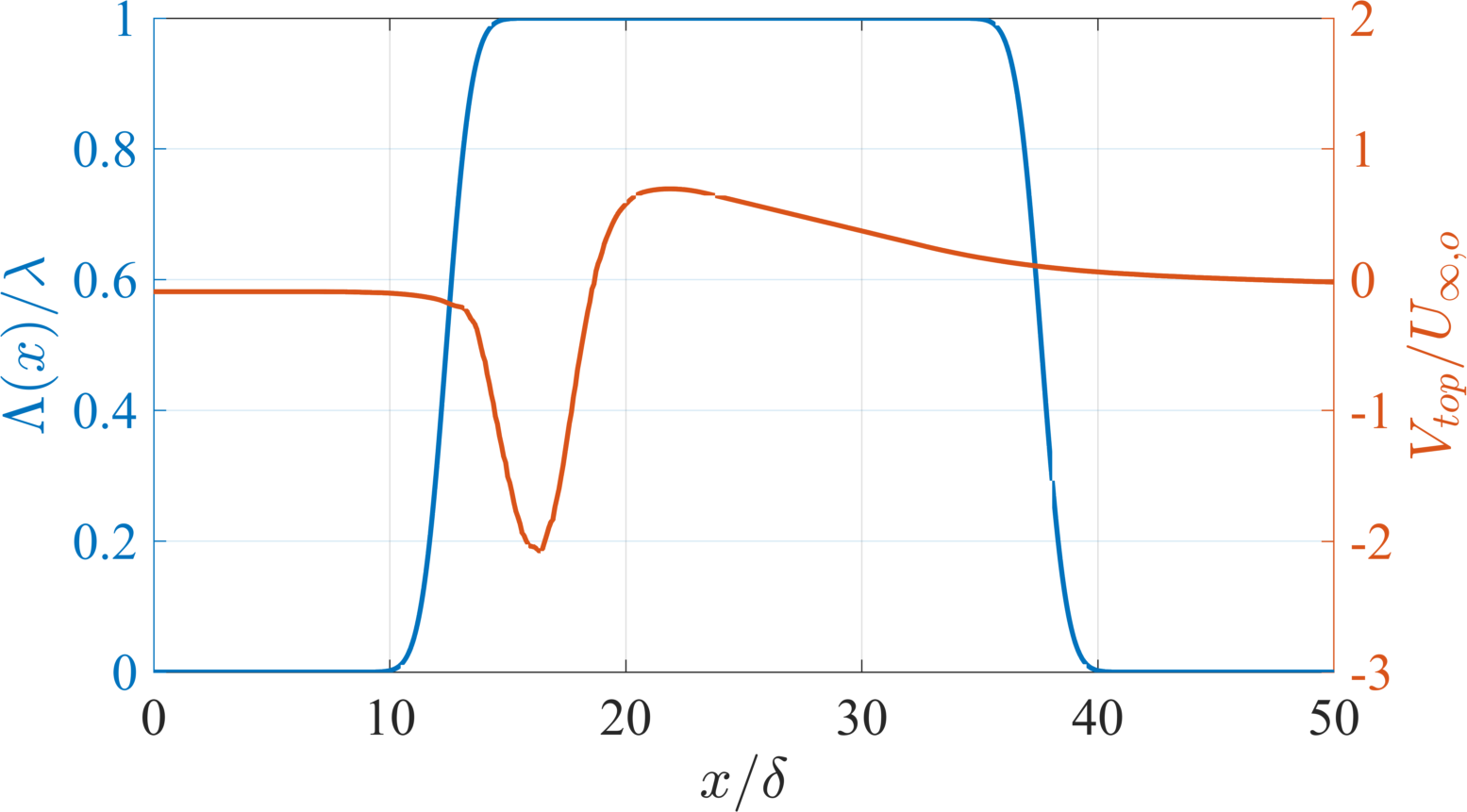} \hspace{2em}    
    \includegraphics[width=0.35\textwidth]{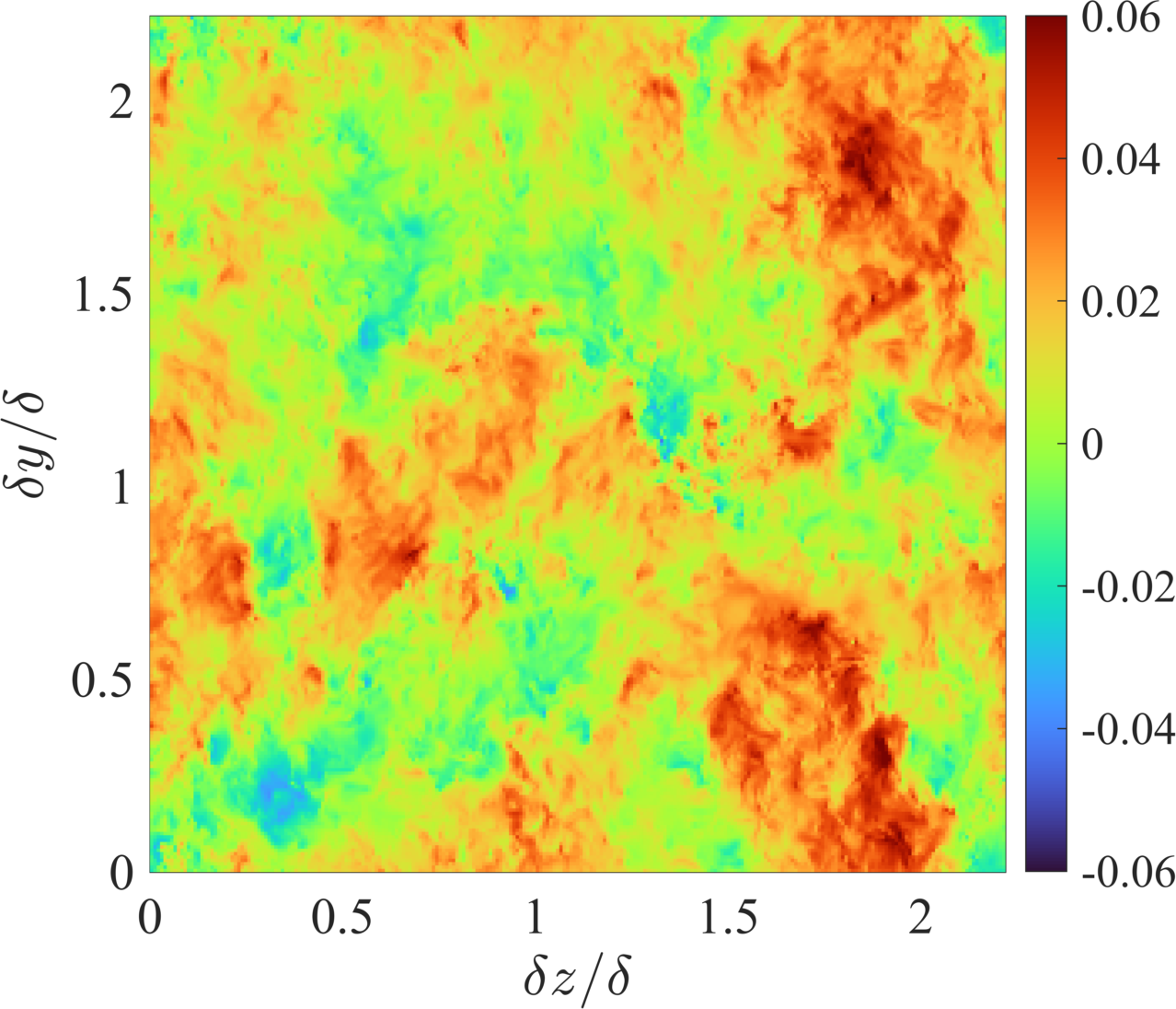}
    \caption{Left, the slip boundary condition profile at the bottom wall and the blowing-suction velocity at the top boundary; right, the homogeneous isotropic turbulence~\cite{JHTDB,CaoC99} added to the inflow. The shown frame is among the total 1024 frames that span about 2.2 time units.}
    \label{fig:slipprof}
\end{figure}

DNS for a laminar boundary layer over a flat plate is performed. Flow separation is introduced by freestream adverse pressure gradient.  
Non-dimensionalized incompressible Naiver-Stokes Equations:
\begin{equation}
    \frac{\partial u_k}{\partial x_k} = 0; \;\;
    \frac{\partial u_i}{\partial t} + \frac{\partial u_i u_k}{\partial x_k} = - \frac{\partial P}{\partial x_i} + \frac{1}{Re} \frac{\partial^2 u_i}{\partial x_k^2}.
    \label{eqn:NS}
\end{equation}
are solved by a well-validated finite difference code that is second-order accurate in time and space~\cite{WuPiomelli18,Wuetal18}.
In the above equations, the indices $(i, k)$ = 1, 2, and 3 correspond to the streamwise ($x$), wall-normal ($y$), and spanwise ($z$) directions. The respective velocity components are $u$, $v$, and $w$. Quantities are normalized by the freestream velocity ($U_{\infty}$) and the boundary layer thickness ($\delta$) at the inflow, leading to a $Re_\delta=U_{\infty} \delta/\nu$ = 455. A blowing-suction velocity profile from Ref. \cite{HosseinverdiF19} is utilized to produce the flow separation. It was measured in a pilot experiment in which an inverted NACA 643-618 airfoil was placed near a flat plate \cite{RadiFasel2010,ChetanFasel2012} (Fig. \ref{fig:slipprof}). 
The streamwise velocity satisfies zero vorticity with the assigned wall-normal velocity profile. 

\begin{figure}
    \centering
    \includegraphics[width=0.85\textwidth]{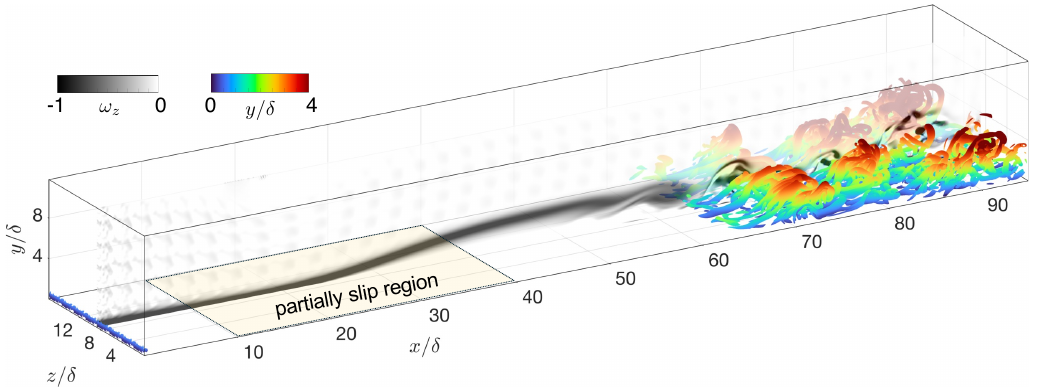}
    \caption{Schematic of the computational setup.}
    \label{fig:config}
\end{figure}

Three cases are studied with various no-slip or partial-slip conditions applied to the bottom wall: Case NS has a no-slip wall over the entire domain. Two partial-slip cases employ a partial-slip condition between $x/\delta$ = 10 and 40. A linear Robin-type slip condition is coupled with the no penetration condition as 
\begin{equation}
    u -  \Lambda(x) \frac{\partial u}{\partial y} =0, v=w=0 \text{, for } y=0.
    \label{eqn:suctionBlowing}
\end{equation}
The first term in the left-hand side of the equation, the slip velocity on the boundary, is determined by a slip length $\Lambda(x)$~\cite{Navier1823}. $\Lambda(x)=0$ represents the no-slip limit, and $\Lambda(x)\rightarrow \infty$ corresponds to a symmetric condition. Its physical meaning can be interpreted by 
\begin{equation}
    u_{y=0} = \frac{\Lambda(x)}{\tau_\nu}, \text{where } \tau_\nu = \frac{\delta_\nu}{u_\tau}
\end{equation}
is the viscous time scale ($u_\tau$ is the friction velocity and $\delta_\nu=\nu/u_\tau$ is the viscous length scale). Therefore, $\Lambda(x)$ describes the distance over which the fluid at the surface travels per unit viscous time scale.  
Instead of the Gaussian profile of $\Lambda(x)$ used by Ref. \cite{Ceccacci1etal22} for their bump, we set the 
\begin{equation}
    \Lambda(x) = \lambda \frac{1+ \text{erf}\left[\left( \frac{L}{2} - 2x_o - \vert x - x_c\vert \right)/x_o\right]}{2},
    \label{eqn:slip}
\end{equation}
where $L=30\delta$ and $x_c=25\delta$ are the length and center of the partial-slip region, respectively. Compared to the Gaussian profile used in Ref. \cite{Ceccacci1etal22}, our profile has a constant $\Lambda(x)$ over the majority of the partial-slip region, i.e., $x/\delta $= [15, 35] (Fig. \ref{fig:slipprof}). $x_o=L/25$ is chosen to give a gradual change between no-slip and the target partial-slip over a $5\delta$-long transition region. $\lambda$=0.1 and 0.2 are used for the two cases, denoted as PS01 and PS02, respectively.

Note that, both the velocity profile at the top boundary and the slip boundary profile at the bottom wall are homogeneous in the spanwise direction. Therefore, the resultant laminar separation bubble (LSB) is statistically two-dimensional in the streamwise-wall normal plane. 
The boundary conditions for the rest of the boundaries of the computational domain are as follows. The periodic boundary condition is applied in the spanwise direction. At the inflow, the Blasius profile of the laminar boundary layer is used. Freestream turbulence is superposed to the mean profile. They are obtained from a database of force homogeneous isotropic turbulence (HIT) at the Johns Hopkins Turbulence Databases (JHTDB)~\cite{JHTDB,CaoC99}. Using Taylor’s hypothesis, one dimension in space of the 3D HIT is converted to time using the free-stream velocity of the boundary layer. The fluctuating velocity components are scaled such that the turbulent intensity ({\it i.e.}, the root-mean-square of the velocity fluctuation) is $Tu=2\%U_{\infty}$ and the integral length scale of the turbulence is about 0.5$\delta$ (Fig. \ref{fig:slipprof}). The outflow is a convective boundary condition.

 The computational domain size 95$\delta$ in the streamwise direction, 11.9$\delta$ in the wall-normal direction, and 17.8$\delta$ in the spanwise direction (see Fig. \ref{fig:config}). They, together with the Reynolds number and the blowing-suction velocity at the top boundary, are chosen to match Ref. \cite{HosseinverdiF19}. The last $5\delta$ of the computational domain is a sponge layer over which the streamwise velocity is adjusted by a body force towards the runtime-averaged streamwise velocity, the spanwise direction, and over $x/\delta=[85,95]$.
 2048$\times$200$\times$256 grid points are used in $x$, $y$ and $z$. The mesh is uniform in $x$ and $z$, while fine near the bottom wall and gradually coarsened away in $y$ with less than 2\% stretching. The maximum grid spacing in wall units is $\Delta x^+ = 2.15$, $\Delta y^+_{(1)} = 0.254$, and $\Delta z^+ =3.23$ upstream of the mean separation point, and less than half of these values for the rest of the computational domain. 
The resolution is finer in all three directions than it is in Ref. \cite{HosseinverdiF19}, ensuring grid convergence. Statistics are collected over 500$\delta/U_{\infty}$ after a statistical steady state is reached in each case. Averaging is conducted over both time and the homogeneous spanwise direction, unless specified otherwise below.

\begin{figure}
\centering
    \includegraphics[width=0.55\textwidth]{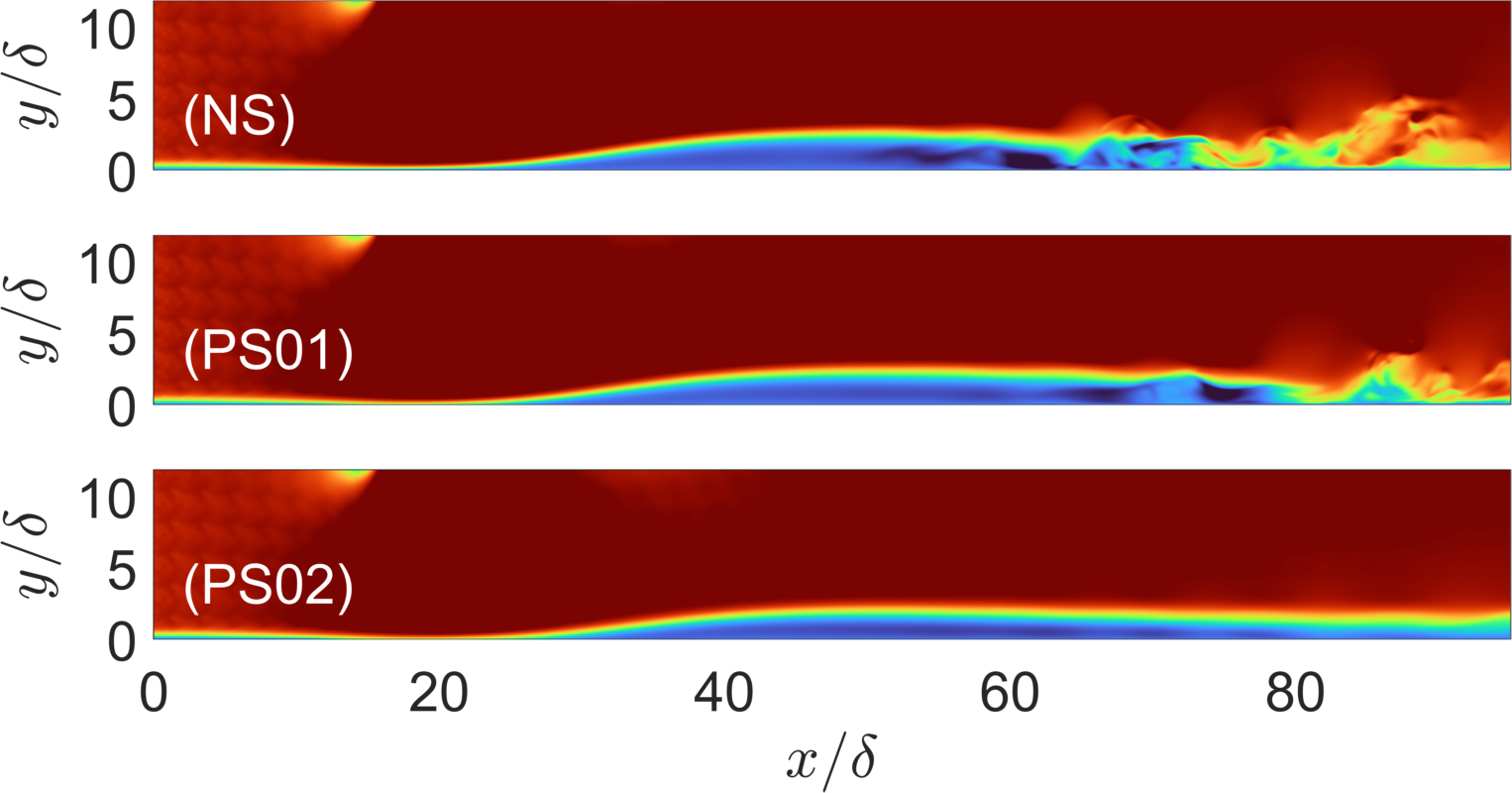}\\
\hspace{3em}\includegraphics[width=0.14\textwidth]{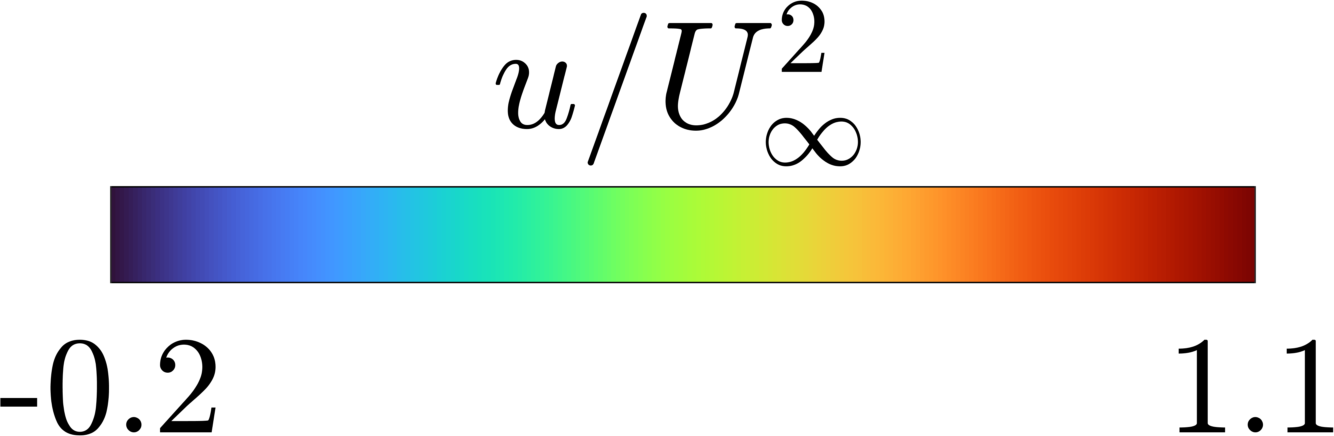}
    \caption{Contours of instantaneous streamwise velocity.}
    \label{fig:instu}
\end{figure}

\begin{figure}
\centering
    \includegraphics[width=0.56\textwidth]{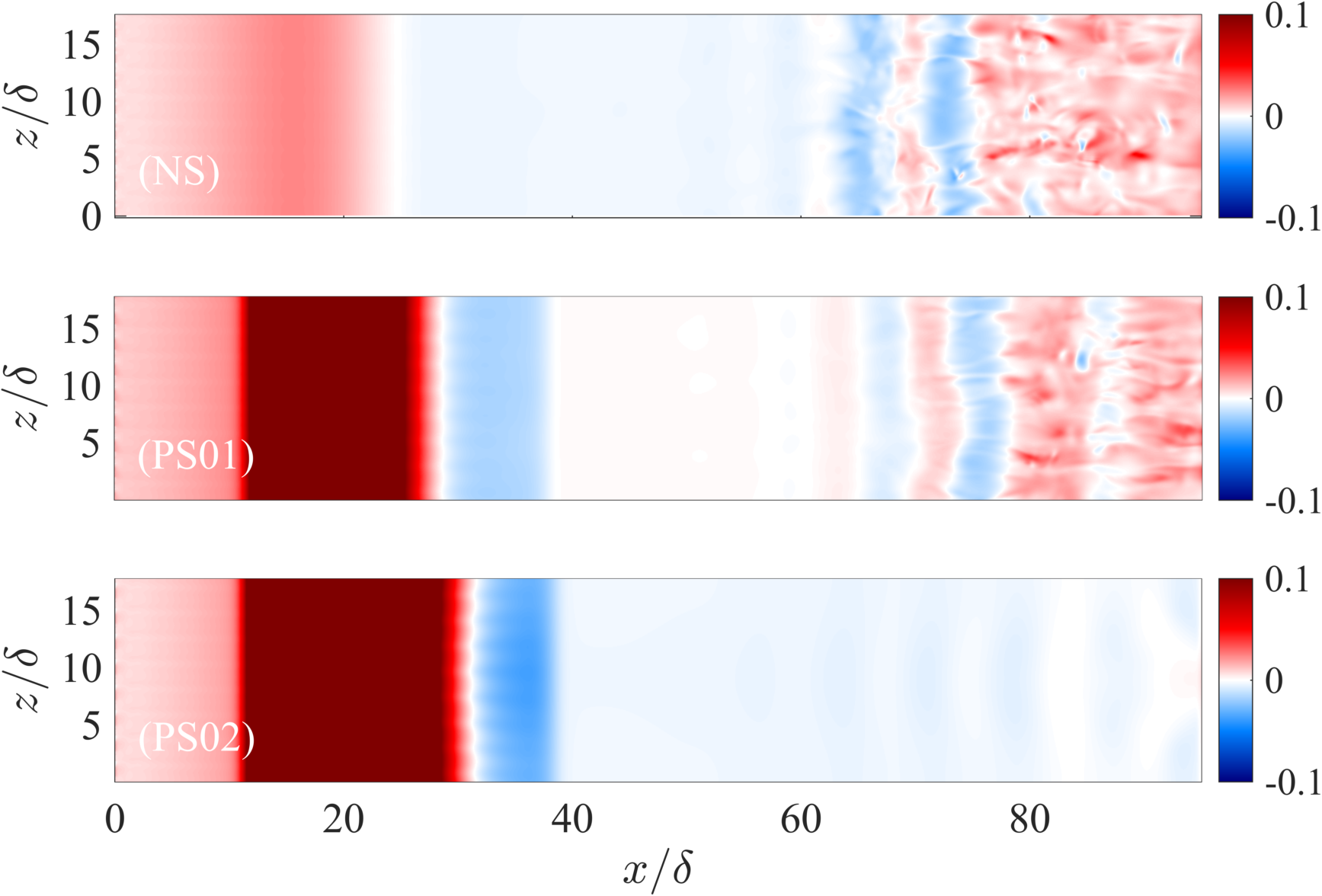}\\
    \caption{Contours of instantaneous streamwise velocity at the first grid point away from the bottom boundary, shown at the same time instants as in figure \ref{fig:instu}. }
    \label{fig:instutop}
\end{figure}

\section{Results and Discussions}
Figures \ref{fig:instu} and \ref{fig:instutop} show the instantaneous streamwise velocity at the mid-span and the first wall-parallel plane away from the bottom boundary, respectively. When the bottom wall is no-slip, the separating shear layer begins to roll up into spanwise vortices around $x/\delta \sim 55$. The roller vortices soon get distorted and break down into hairpin vortices (refer to Fig. \ref{fig:config}). When the partial-slip boundary condition is applied in Cases PS01 and PS02, the shed-off vortices are weaker and more two-dimensional than in the no-slip case. They can be identified by the footprints on the bottom boundary. In Case PS01, the formation of the roller vortices is delayed to around $x/\delta \sim 60$ and the three-dimensionalization can be observed only near the outflow. At the higher slip velocity, the spanwise roller vortices are weak and no vortex breakdown is observed within the computational domain. The vortex shedding frequency, indicated by the spatial spacing between the roller vortices in both figures, does not vary significantly. Time series analysis in this region will be performed in the full paper to further justify the shedding frequency.

\begin{figure}
\centering
    \includegraphics[width=0.65\textwidth]{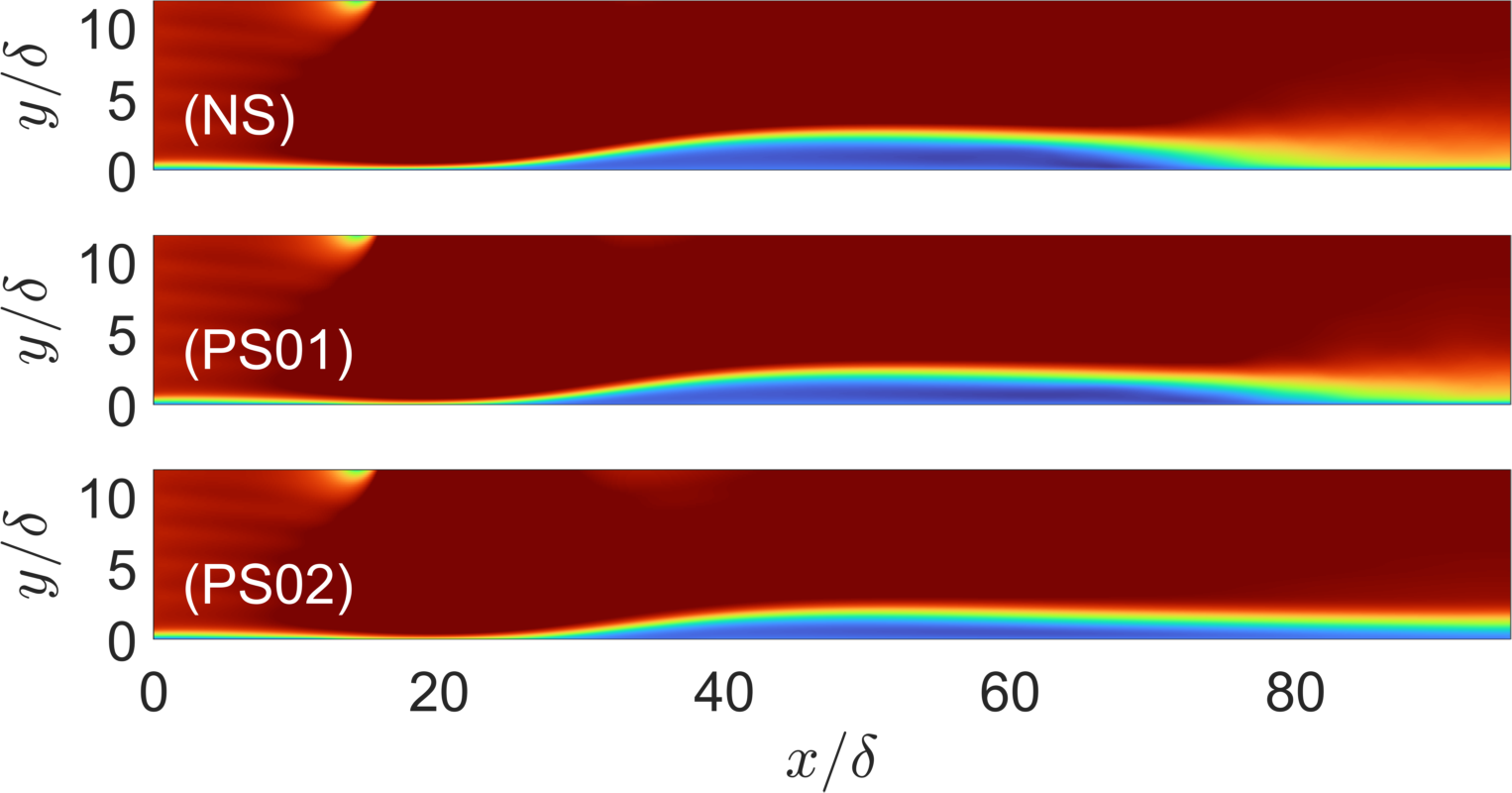}\\    \hspace{3em}\includegraphics[width=0.18\textwidth]{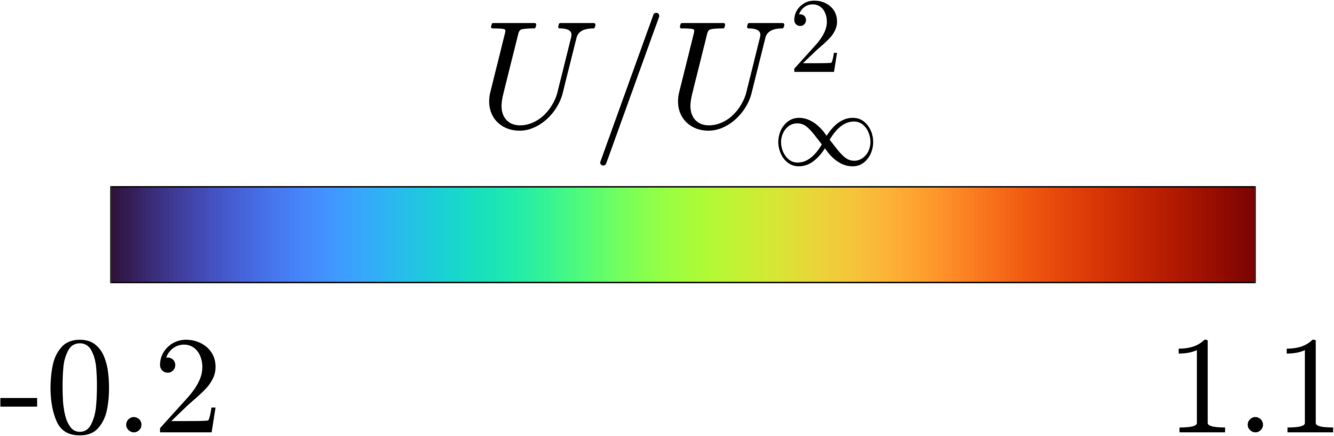}
    \caption{Contours of the mean streamwise velocity.}
    \label{fig:meanu}
\end{figure}

\begin{figure}
\centering
    \includegraphics[width=0.9\textwidth]{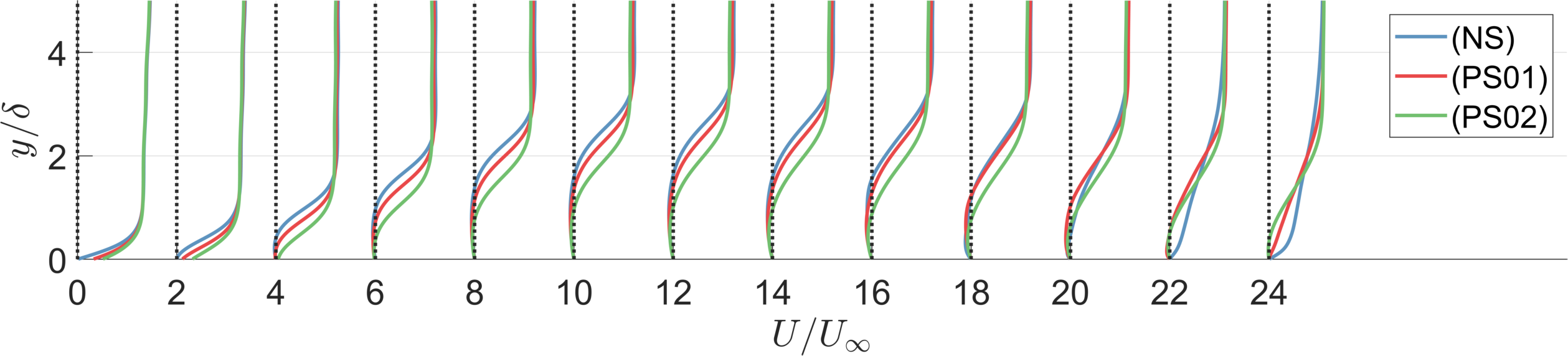}\\
    \caption{Comparison of mean streamwise velocity near the bottom wall at selected $x$ locations. From left to right, $x/\delta = 20, 25, 30, 35, \dots, 80$. Each profile is shifted to the right by 2 units for clarity.}
    \label{fig:uprofile}
\end{figure}

The mean flow is compared in figures \ref{fig:meanu} and \ref{fig:uprofile}. The mean separation bubble length increases with the slip velocity and the thickness of the separation bubble decreases with it. Specifically, the onset of separation is delayed by the slippery wall, agreeing with the previous literature~\cite{LegendreLM09,Soorajetal20,MURALIDHARetal11}. Referring to the mean skin friction profile, it can be identified that the mean separation occurs at $x/\delta$ = 25.9, 28.6 (+2.7$\delta$), and 32.0 (+6.1$\delta$) for Cases NS, PS01, and PS02, respectively.  Compared with the mean separation point, the mean reattachment is further postponed due to the weakened vortex: Case PS01 and PS02 reattach 5.0$\delta$ and 13.8$\delta$ further downstream than Case NS. Therefore, the recirculation region is increased by the slipping wall. 

Note that the wall shear stress is obtained based on Eq. (\ref{eqn:suctionBlowing}) using the mean velocity in the vicinity of the wall.
The $C_f$ profiles show several other features that represent the dynamics of the separating shear layer: 1) the wall shear stress first decreases at the beginning of the partial-slip region, then increases as the slipping level plateaus. The occurrence of this increase also coincides with the favorable pressure gradient (FPG) introduced by the blowing part of the velocity at the top boundary (refer to Fig. \ref{fig:slipprof}). Then, around the same location where $C_f$ starts to decrease in the no-slip case, the $C_f$ for the two partial-slip cases also declines as the flow approaches the separation point. The same trend can be observed for the slip velocity. It indicates a rich interaction between the change of velocity deficit due to the slipping surface and the modulation due to pressure gradients. The FPG and the slipping boundary both accelerate the near-wall flow, while the APG opposes it.
%
2) The reversed flow is comparable between the three cases within the majority of the recirculation region. It means that there is little interaction between the flow at the separation and reattachment points. The slipping boundary has little effect when the near-wall flow is close to stationary inside the separation bubble. 

\begin{figure}
\centering
    \includegraphics[width=0.56\textwidth]{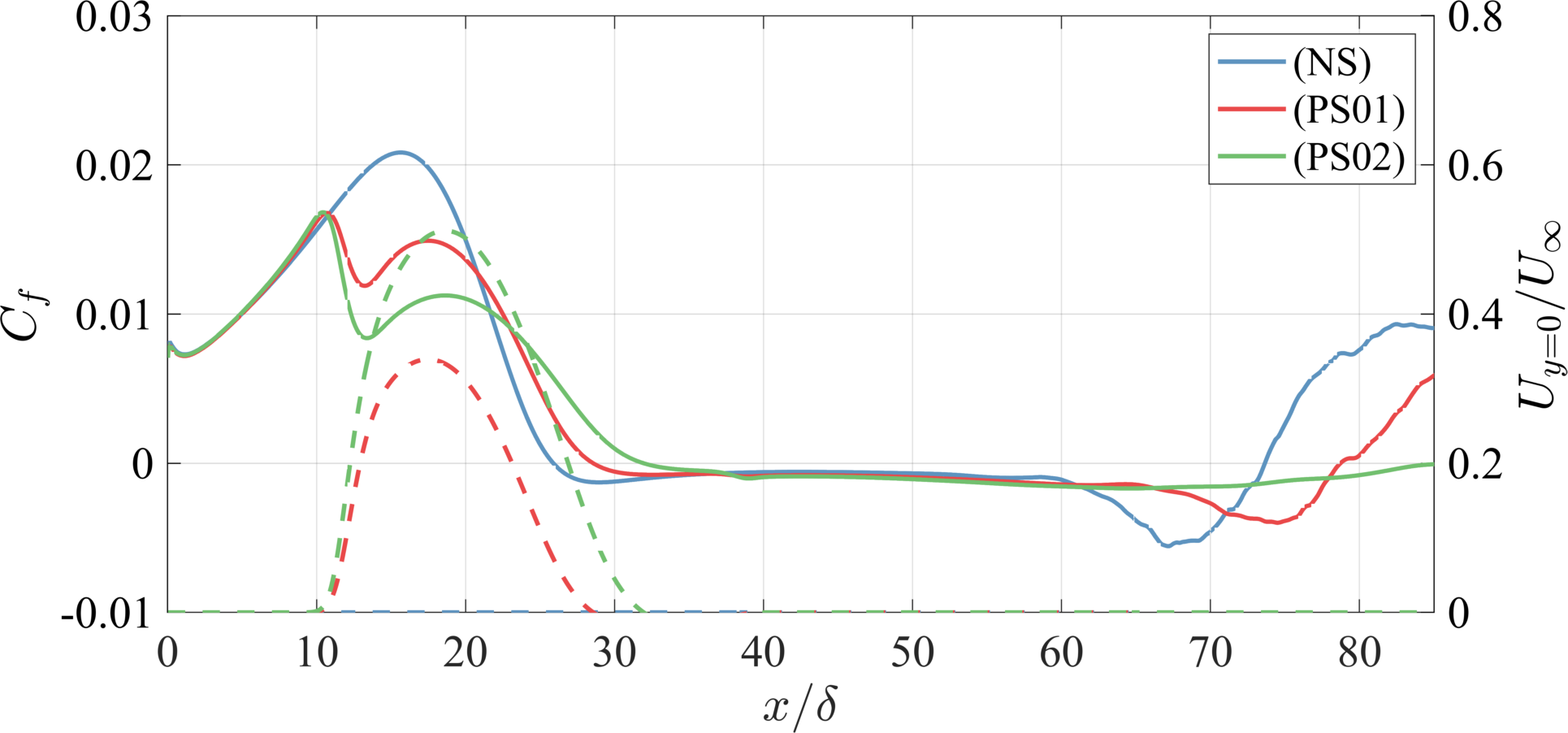}\\
    \caption{Profiles of the mean skin friction coefficient and velocity at the bottom boundary. Solid, $C_f$ (left axis); dashed, mean velocity at the bottom boundary (right axis).}
    \label{fig:cf}
\end{figure}

\begin{figure}
\centering
    \includegraphics[width=0.8\textwidth]{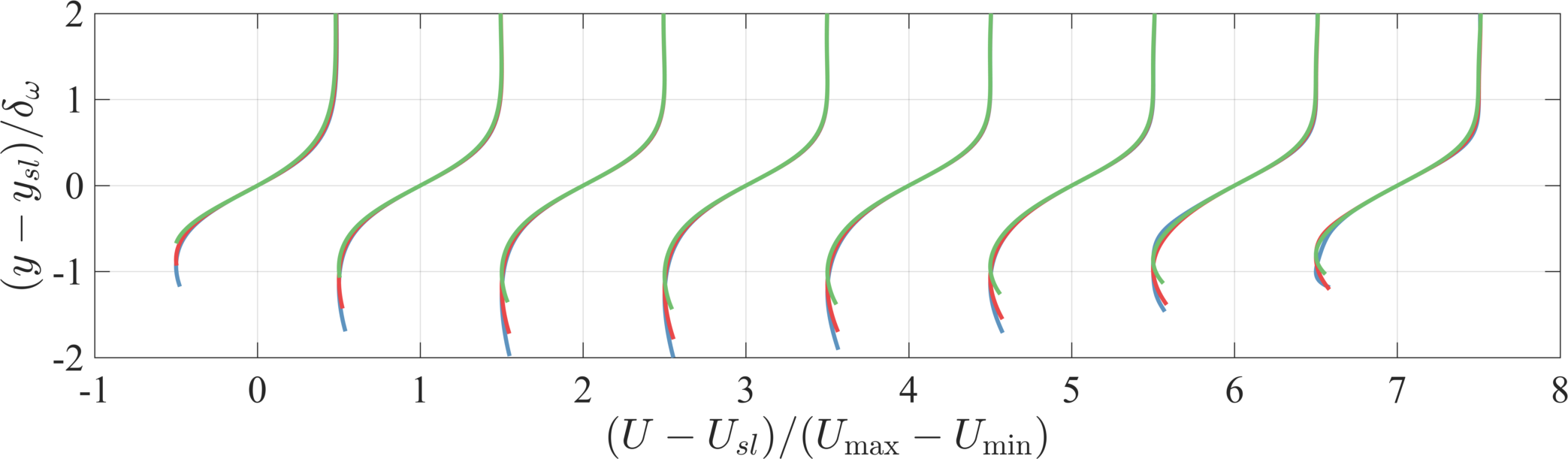}\\
    \caption{Mean velocity scaled by shear layer scaling at selected $x$ locations. From left to right, $x/\delta = 30, 35, 40, \dots, 65$. Each profile is shifted to the right by 1 unit for clarity. Line styles are the same as figure \ref{fig:uprofile}.}
    \label{fig:u_norm}
\end{figure}

\begin{figure}
\centering
    \includegraphics[width=0.9\textwidth]{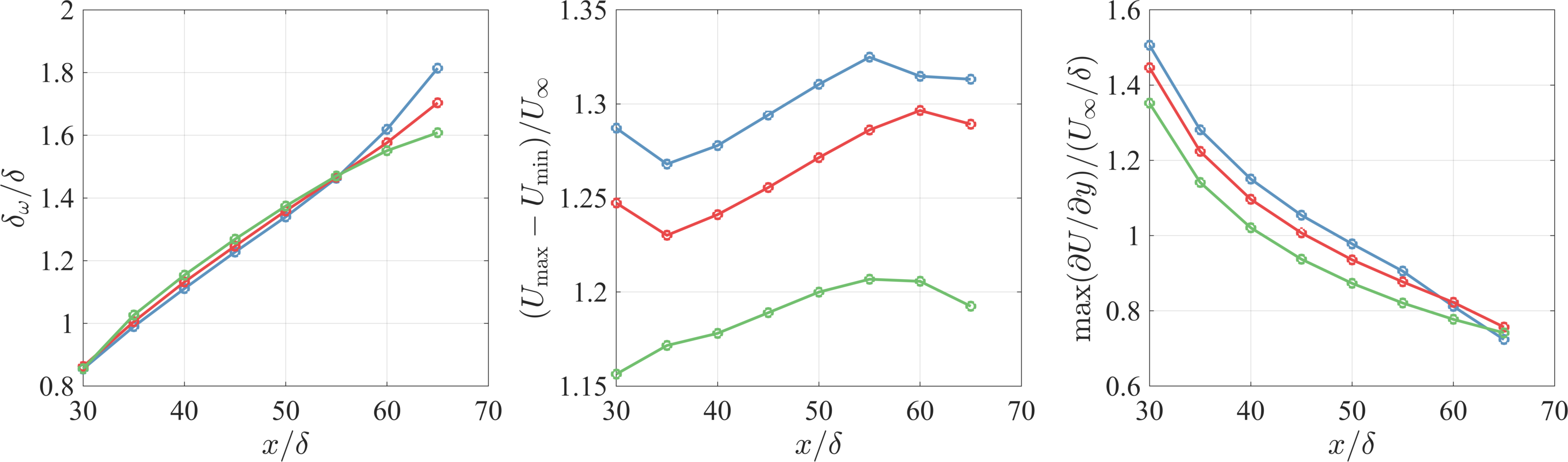}\\
    \caption{Streamwise variations of the shear layer scaling parameters. Line styles are the same as figure \ref{fig:uprofile}. }
    \label{fig:trend}
\end{figure}

The reduced velocity deficit by the partial-slip boundary later appears in the separating shear layer as the latter detaches from the surface. Figure \ref{fig:uprofile} compares the streamwise velocity at selected streamwise locations. It can be seen that at a certain distance away from the bottom wall, the velocity in the forward-flow side of the separating shear layer increases with the slipping velocity. As a result, the separating shear layer departs less from the surface. To examine the degree of similarity to a plane mixing layer, we have examined the profile of the separated shear layer based on scaling parameters of plane mixing layers:
\begin{equation}
\delta_\omega =(U_\text{max}-U_\text{min})/(\partial U/\partial y)_\text{max}
\end{equation}
defines the vorticity thickness where $U_\text{max}$ and $U_\text{min}$ are the maximum and minimum mean velocities in the two sides of the separated shear layer \cite{Abe19,LiepmannL47}. Besides, the shear layer velocity is defined as 
\begin{equation}
U_{sl} = (U_\text{max} + U_\text{min})/2
\end{equation}
and $y_{sl}$ is where it occurs in the wall-normal direction. The velocity scaled by these quantities, and the evolution of these scaling parameters in the three cases, are compared in figures \ref{fig:u_norm} and \ref{fig:trend}. The self-similarity of all velocity profiles is clear except for deep in the recirculation region. The separating shear layer roughly maintains the same thickness between the three cases up to $x/\delta=55$. The velocity difference on the two sides of the separating shear layer decreases with the slipping velocity. Therefore, less shear is present on average over the entire separating shear layer as the slipping velocity increases. The maximum shear is also reduced, highlighting the reason behind the reduced vortex shedding. 
Downstream of $x/\delta=55$, the maximum shear becomes comparable between cases. The velocity difference between the two sides of the separating shear starts to plateau or decrease. This is where the shear layer begins to roll up into vortices.

\section{Conclusion}
DNS of a laminar separating boundary layer induced by freestream pressure gradient is performed. With a partial-slip region upstream of the separation region, the separation bubble becomes longer and thinner. Both the onset of separation and the reattachment are delayed. Formation and breakdown of roller vortices in the separating shear layer are mitigated by the slipping surface. The self-similarity solution of the plane mixing layer is still valid as the laminar boundary layer with less velocity deficit detaches from the surface. The scaled overall and maximum velocity gradient across the shear layer increases with the slipping velocity.
The results highlight the possible impact of strengthened hydrophobicity on the aerodynamics of wings, turbines, and nozzles that are subjected to adverse pressure gradients and potential flow separation. The current configuration appears to show that a partial-slip region upstream of the separation is beneficial for reducing flow separation and maintaining the laminar flow state. 

New open questions are motivated by the results, including the sensitivity with respect to the position and extent of the partial-slip region, vortex dynamics if the reattachment region is also partial-slip, and how turbulent boundary layers respond to a partial-slip boundary. The full paper will provide more details regarding the parametric study with various freestream turbulence levels, slip region sizes, and Reynolds numbers.

\section*{Acknowledgments}
The authors acknowledge the support of the San Diago Supercomputing Center (SDSC) for providing computational resources on the Expanse cluster. Wu acknowledges the support from NSF grant CBET-2039433, monitored by Dr. R. Joslin. B.S. acknowledges support from NSF GRFP Award 2235036.  

\bibliography{main}

\end{document}